\documentstyle[twocolumn]{article}
\def\section#1{\bigskip\centerline{\bf #1}\medskip}
\def\subsection#1{\medskip\leftline{\small\bf #1}\smallskip}
\def\acknowledgement#1{\medskip\centerline{\small\bf Acknowledgements}
		       \smallskip\noindent{\small #1}}

\begin{document}

\title{\rightline{\small IISc-CTS-4/01}\vspace*{-0.4truecm}
       \rightline{\small quant-ph/0102034}
{\Large\bf Testing quantum dynamics in genetic information processing}}

\author{\normalsize APOORVA PATEL\thanks{E-mail: adpatel@cts.iisc.ernet.in}\\
       {\normalsize\it CTS and SERC, Indian Institute of Science,
       Bangalore-560012, India}}

\date{{\normalsize\bf Abstract}\\
\medskip\leftline{\small
Does quantum dynamics play a role in DNA replication? \hfil
What type of tests would reveal that? \hfil Some statistical checks}
\vspace*{-0.1truecm}\leftline{\small
that distinguish classical and quantum dynamics in DNA replication
are proposed.}
\medskip\leftline{{\normalsize\bf Keywords.}\hfill\normalsize
assembly; computation; database search; DNA replication; genetic
information; nucleotide base;}
\vspace*{-0.1truecm}\leftline{\normalsize
polymerase enzyme; quantum coherence; quantum mechanics; quantum
superposition.}
\medskip\leftline{\small
[Patel A. 2001 Testing quantum dynamics in genetic information processing.
J. Genet. {\bf 80}, 39-43]}
\medskip\hrule}

\maketitle

\section{Genetic information processing}

\noindent
Darwinian evolution---survival of the fittest---is an optimisation problem.
Its foundation is the observation that living organisms have adapted to
their environment, and have exploited the available material resources
and the physical laws governing them to the best of their capability.
The environmental influences are usually local and so are the adaptations.
So the meaning of ``best of their capability'', is different in different
situations, and it is an interesting exercise to figure out the physical
principles involved in the optimisations.

The genetic code is a feature of living organisms that has remained virtually
unchanged from the ancient bacteria to modern human beings. Its purpose is to
convey the hereditary genetic information from one generation to the next.
The task is carried out by the long chains of DNA molecules residing in the
nuclei of the cells, which faithfully replicate themselves each time the cell
divides. In the early studies of molecular biology, it was proposed that the
genetic code was a frozen accident (Crick 1968).
More recent investigations have shown that some parts of the genetic code
have flipped back and forth independently in completely different organisms
(Knight {\it et al.} 2001).
This leads credence to the belief that the genetic code is evolvable and
optimisation criteria have been involved in its design.

What are then the optimisation criteria involved in genetic information
processing? The task involved is an unsorted assembly operation: pick up the
building blocks from their random mixture in the environment, check whether
they have the desired properties or not and then join them together in the
requisite order. This is a well-defined computational problem, and it is
reasonable to suppose that optimisation corresponds to carrying out the task
as quickly as possible and without errors. The solution to this problem
depends on the type of dynamics used to implement the algorithm; in particular
the optimal solution is different for classical and quantum dynamics.

I have proposed that assumption of quantum dynamics in the assembly operation
would naturally explain why the genetic information is organised in a base-4
language, and not in the base-2 language of our classical digital computers
(Patel 2000b).
My study of the molecular structure of DNA showed that it indeed has the
capability to implement the quantum algorithm (Patel 2000a).
The question that remains is: does DNA replication actually use the quantum
algorithm, or is the existence of both software and hardware features an
accident?

Of course, the structure of DNA came into existence billions of years ago,
and it could be that what was relevant when life arose is not relevant now;
i.e. the observed features are hanging around from a bygone era like the
human appendix. Unfortunately, we cannot recreate the conditions in which
life originated, and so cannot test whether quantum dynamics played a part
in it or not. What we can test is whether quantum dynamics plays a role in
DNA replication today or not. Direct observation of the assembly operation
as it happens would be an ideal test, but our technology has not yet
progressed to that level. We have to rely on statistical tests to infer the
dynamical process.

Here I point out some statistical tests, that can differentiate between
classical and quantum dynamics in DNA replication. They crucially depend
on our ability to construct ``designer DNA'' and use it in the replication
process. The following section describes tests based on DNA replication rates,
while the subsequent section describes tests based on the DNA structure.

\section{Replication rate checks}

\noindent
First let us go through the mathematical description of the assembly
algorithm. That clearly brings out the distinction between classical and
quantum dynamics, and also explains the criterion for optimisation.

\subsection{Classical unsorted assembly process}

\noindent
The assembly process is a variation of the search process, where the objective
is not to locate the desired item in a database but to assemble it using the
available building blocks. The building blocks can be of many different types
distinguishable from each other by certain properties. Let $a$ be the number
of types of building blocks and $n$ be the length of the chain to be assembled.
Then the total number of items that can be assembled is $N=a^n$. $N$ is the
measure of information contained in the chain, and $a\ge2$ is necessary to be
able to convey worthwhile information.

We consider the situation where building blocks are available in large
numbers for each type. Also they are available not in a particular order,
but as a random ensemble of all possibilities. Let $t_d$ be the time required
to pick up a building block from the random ensemble, and $t_a$ be the time
required to add each building block to the growing chain. Since the ensemble
of the building blocks is random, the building block that is picked up may
or may not be the one desired for adding to the growing chain, and therefore
$t_d\ne t_a$. Certain property checks decide whether the building block
picked up is the desired one or not. If it is the desired one, it is added to
the growing chain. If it is not the desired one, it is discarded and another
one is picked up from the random ensemble. Again its properties are checked,
and the trial and error process continues till the desired building block is
found. We also assume that the property check process has no memory, i.e.
there is no correlation between a building block that is discarded and the
new one that is picked up next. The ensemble of building blocks is thus truly
random; what is discarded once can be picked up immediately again. In this
situation, the probability that the picked building block is the desired one
is $1/a$.

The total time required to classically assemble the complete chain is ($N$
is chosen to be sufficiently large so that round-off errors can be ignored):
\begin{equation}
T_c(a) = t_a n = t_d a n = t_d a \log_a N = t_d a (\ln N / \ln a) ~.
\end{equation}
The optimisation criterion for the assembly process is to minimise $T_c(a)$
by varying $a$, while holding $t_d$ and $N$ fixed. Holding $N$ fixed means
that the information to be conveyed is fixed. Holding $t_d$ fixed means that
the total number of all the available building blocks in the random ensemble
is fixed; the concentration of a particular type of building block will be
proportional to $1/a$. This minimisation criterion decides the number of
types of building blocks required for the quickest processing of a fixed
amount of information. The best value for $a$ obtained by minimising
$T_c(a)$ is not an integer, $a_{\rm min} = e = 2.71828\ldots$. The integer
solutions close to the lowest value of $T_c(a)$ are:
\begin{eqnarray}
T_c(a=2) &= {2 \over \ln 2} t_d \ln N \approx 2.8854 t_d \ln N ~, \cr
T_c(a=3) &= {3 \over \ln 3} t_d \ln N \approx 2.7307 t_d \ln N ~, \cr
T_c(a=4) &= {4 \over \ln 4} t_d \ln N \approx 2.8854 t_d \ln N ~.
\end{eqnarray}
For larger $a$, $T_c(a)$ increases monotonically. Thus the best choice is
$a=3$, while $a=2$ and $a=4$ would serve equally well as the next best
possibility.

\subsection{Classical DNA replication rates}

\noindent
In the particular case of DNA replication, the building blocks are the
nucleotide bases, and property checks are performed by molecular bonds
involved in the base-pairing. $t_d$ is the sum of the time required for a
base to diffuse from the environment into the enzyme cavity and the time
required for making property checks. The replication time per base-pair,
$t_a$, can be easily measured in experiments. Measurement of $t_d$ would
require observing the accept/reject process for a picked base, which is
not yet directly observable. The analysis of the previous section predicts
specific DNA replication rates depending on the value of $a$.

Arbitrary choices of $a$ are not possible with the naturally occurring
molecules, yet situations can be selected to correspond to some of the
values of $a$:\\
$\bullet$ Genetic DNA of living organisms has $a=4$, consisting of two
complementary base pairs, A-T and C-G.\\
$\bullet$ DNA constructed with one of the complementary base pairs and a
single base on one strand would correspond to $a=3$. The single base should
either pair with itself or the environment should contain its complementary
base. (For example, the strand to be replicated could be made of only A,C,G
while the environment has only T,C,G.)\\
$\bullet$ DNA constructed with only one of the complementary base pairs would
correspond to $a=2$. $A-T$ rich and $C-G$ rich DNA already exist in nature,
and they can be modified to eliminate one of the complementary base pairs.\\
$\bullet$ DNA constructed with a single base (as in case of $a=3$) on one
strand would correspond to $a=1$. Such DNA will not carry information,
but its replication rate can still be measured.

The classical base-pairing rates as a function of $a$ are predicted to be
\begin{equation}
R_c(a) = n/T_c(a) = (1/a) t_d^{-1} ~.
\end{equation}
The unknown $t_d$ can be eliminated by forming ratios of replication rates.
Such ratios can be determined more accurately than the absolute rates, since
they cancel unknown systematic errors in $t_d$. For example, it is unlikely
that every time the correct nucleotide base diffuses in the enzyme cavity,
it will bind with the intact strand. Wasted encounters with the correct
nucleotide base can be accounted for by just rescaling $t_d$.

Another possibility for a systematic effect is that the times for $A-T$ and
$C-G$ base-pairing are different, since they involve different free energies.
This can be tested by comparing the replication rate of $a=2$ $A-T$ strands
with that for $a=2$ $C-G$ strands. If there is a systematic effect, the
replication rates can be renormalised to reflect the proportion of each
base-pair.

Thus, by constructing DNAs corresponding to different values of $a$ and
comparing their replication rates, we can experimentally test whether DNA
replication follows classical dynamics or not.

\subsection{Quantum DNA replication rates}

\noindent
The quantum search algorithm works faster than its classical analogue,
by using clever interference of a superposition of states (instead of
individual states) (Grover 1996).
The algorithm can be described as a unitary evolution of a set of states
(one state corresponding to each possibility) in the quantum Hilbert space.
At the beginning, any starting state is converted into a uniform superposition
of all the possibilities. Then two steps are executed iteratively:
(i) the amplitude of the desired state is flipped in sign by the property
checks, and (ii) all the amplitudes are reflected about their average value
by an overrelaxation procedure. These iterations amplify the amplitude of
the desired state, and diminish the amplitudes of the remaining states.
By stopping the iterations at the right moment, the desired state is
selected with high probability.  The number of iterations required, $Q$,
is related to the number of items to be distinguished, $a$, by
\begin{equation}
(2Q+1) \sin^{-1} (1/\sqrt{a}) = \pi/2 ~.
\end{equation}
For a given value of $a$, the solution for $Q$ is not necessarily an integer
(the exception is $Q=1$ for $a=4$). For non-integral $Q$, the number of
iterations is to be interpreted as the integer closest to and larger than $Q$.

Let $t_r$ be the time required to add one building block to the chain using
the quantum algorithm. In addition to the time required to pick up a building
block from the environment and performing property checks, it includes the
relaxation time required for creating the necessary superposition state.
Thus $t_r$ is expected to be larger than the time $t_d$ of the classical
algorithm. The total time required to assemble the complete chain by the
quantum algorithm is proportional to the number of iterations, and is not
a smooth function of $a$. The lowest values are given by:
\begin{eqnarray}
T_q(a\le4) &= t_r n = t_r (\ln N / \ln a) ~, \cr
T_q(5\le a\le10) &= 2 t_r n =  2 t_r (\ln N / \ln a) ~.
\end{eqnarray}
The minimum value of $T_q(a)$ is obtained for $a=4$. For the DNAs constructed
to correspond to different values of $a$, as described above, the quantum
base-pairing rate is
\begin{equation}
R_q(a\le4) = n/T_q(a) = t_r^{-1} ~.
\end{equation}
This rate is independent of $a$---distinctly different from the classical
prediction. Therefore, experiments measuring DNA replication rates with
different values of $a$ can distinguish classical and quantum dynamics.

In reality, the quantum algorithm is unlikely to work with 100\% efficiency.
Disturbances from the environment, i.e. decoherence effects, cannot be wished
away. It should be noted that, with different quantum states represented by
different physical objects, the quantum algorithm needs only superposition
and interference, and does not need the far more fragile property of quantum
entanglement (Lloyd 2000).
Whenever the quantum algorithm is disrupted by decoherence effects, the
replication process can still continue with multiple attempts as in the
classical case. Even an imperfect quantum algorithm will beat the best
classical alternative, if its success probability obeys
\begin{equation}
p > T_q(a=4)/T_c(a=3) = 0.264 t_r/t_d ~.
\end{equation}
This is a substantial tolerance margin for the quantum algorithm, provided
that $t_r$ is not significantly larger than $t_d$. Biological systems are
known to sense small differences in population growth rates (e.g. growth
of type $dP/dt = r P$). Even an advantage of a fraction of a percent is
sufficient for one species to overwhelm another, provided there is long
enough time for evolution. Nonetheless, it should be noted that an imperfect
quantum algorithm will be harder to verify, because its properties will lie
somewhere in between the classical and the quantum predictions.

\section{Structural checks}

\noindent
Quantum superposition of states is a crucial ingredient of any quantum
algorithm that outperforms its classical counterpart. Processing of a
linear superposition of states is a type of parallel processing which can
speed up certain algorithms without requiring additional resources. The
superposition of states has to be created from the classical state used
to initialise the algorithm, and thereafter quantum coherence has to be
maintained during the course of the algorithm.

The quantum assembly algorithm requires that a uniform superposition of
all possibilities be created from any classical starting state. This is
possible only if the quantum dynamics provides transition matrix elements
amongst the classical states. (The transition matrix elements produce a
mixture of states in a random classical environment, but they produce
superposed states in a coherent quantum environment. I have hypothesised
that polymerase enzymes help carry out these transition processes
(Patel 2000a).)
The classical states are nucleotide bases with different chemical
compositions, so the quantum transition processes must involve exchange
of chemical groups. Such exchange of chemical groups can be detected by
isotopic tagging; radioactive processes or NMR techniques can track the
shift of isotopes from one location to another.

The nucleotide bases consist of a large group of atoms common to all of them,
and a small group of atoms which tell them apart. The superposition state
can be created by many different types of quantum transition processes. But
any such quantum transition process will need to exchange the small group of
atoms between different nucleotide bases or between a nucleotide base and the
polymerase enzyme. Consider a situation where one of the nucleotide bases in
the environment has some of the atoms in both the large and small group of
atoms replaced by different isotopes. If there are no exchanges of chemical
groups, then the two isotopically tagged groups of atoms will be always found
together. On the other hand, if the exchanges of chemical groups do take place,
then the two isotopically tagged groups will get separated, at least in some
of the molecules. In such a case, after the replication is complete, the
isotopically tagged small group of atoms will be found either in another
nucleotide base (most likely incorporated in the newly constructed DNA strand)
or in the polymerase enzyme. This can be experimentally detected.

Another point worth noting is that if the polymerase enzyme plays an active
role in the exchange of chemical groups, then it is likely to have all the
necessary chemical groups in close vicinity of its active sites, so that the
exchange may take place smoothly. Detailed investigations of the polymerase
enzyme structure can confirm this.

\section{Summary}

\noindent
To take quantum processing of genetic information beyond an explanation,
its underlying hypotheses must be checked against falsifiable predictions.
It is not yet technologically feasible to observe the dynamics of DNA
replication, so I have pointed out some statistical tests that predict
different results for classical and quantum dynamics. The predictions
are based on the dynamical steps involved in the algorithm, and so they
indirectly verify the dynamical process.

The tests are of two types: some measure the replication rates and some
check the structure of the participating molecules. The replication rate
tests require the number of building blocks in the DNA strand to be variable.
The protein coding sections of naturally occurring DNA have all the four
nucleotide bases. On the other hand, stretches of DNA strands made up of
fewer bases are known to exist in the non-coding sections of DNA, and it
should be possible to extract strands having one, two or three building
blocks from them. The structural checks require introduction of isotopically
tagged nucleotide bases in the environment, and then tracking the various
isotopes to their final destinations. Such tests have been carried out in
many chemical processes.

These tests use only naturally available molecules, and so they check the
optimality of the DNA replication process only in a limited sense.
In particular, the building blocks of DNA are varied, but the enzymes are
left unchanged. The presence of the enzymes is a must for the DNA replication
process; random molecular collisions do not replicate DNA. It is quite likely
that properties of the enzymes and the building blocks are correlated, which
implies that the optimal replication algorithm should be searched for by
simultaneously changing the enzymes as well as the building blocks. I have no
clue regarding how to find the optimal enzyme. In this sense, the suggested
tests do not really confirm the optimality of the DNA replication process,
although the quantum assembly algorithm was proposed on the basis of
optimality.

Finally, I want to point out that I have used the language of DNA replication
for describing the tests, but the tests can equally well be carried out for
the mRNA transcription process. The base-pairing algorithmic steps are the
same in both cases, even though there are chemical differences in the building
blocks and the polymerase enzymes used. In fact, the tests should be easier
to implement for the mRNA transcription process, since it involves only one
DNA strand (in contrast to both the strands being simultaneously involved in
the DNA replication process).

\acknowledgement{\small
I am grateful to all those who exhorted me to make quantitative predictions
based on my proposal that quantum dynamics plays an important role in
genetic information processing. I particularly thank L. Grover, N.V. Joshi,
V. Nanjundiah, L. Quidenus, H. Sharatchandra, U. Varshney and M.A. Viswamitra
for their encouragement and feedback.
}

\bigskip
\centerline{\bf References}
\medskip

{\small
\noindent
Crick F.H.C. Crick 1968 The origin of the genetic code.
{\it J. Mol. Biol.} {\bf 38}, 367-379.

\noindent
Grover L. 1996 A fast quantum mechanical algorithm for database search.
In proceedings of the 28th annual ACM symposium on theory of computing,
Philadelphia, pp.212-219 [quant-ph/9605043].

\noindent
Knight R.D., Freeland S.J. and Landweber L.F. 2001
Rewiring the keyboard: Evolvability of the genetic code.
{\it Nature Reviews Genetics} {\bf 2}, 49-58.

\noindent
Lloyd S. 2000 Quantum search without entanglement.
{\it Phys. Rev.} {\bf A61}, 010301(R), pp.1-4 [quant-ph/9903057].

\noindent
Patel A. 2000a Quantum algorithms and the genetic code. In proceedings of
the winter institute on foundations of quantum theory and quantum optics,
Calcutta, {\it Pram{\=a}{\d n}a - J. Phys.} {\bf 56}, 367-381
[quant-ph/0002037].

\noindent
Patel A. 2000b Quantum database search can do without sorting.
{\it Phys. Rev.} {\bf A} (in press) [quant-ph/0012149].
}

\end{document}